\documentclass[twocolumn,prb,aps,showpacs,floatfix,superscriptaddress]{revtex4}
\usepackage{amsmath}
\usepackage{amssymb}
\usepackage[dvips]{graphicx}
\usepackage{dcolumn}
\usepackage{bm}
\usepackage{ifpdf}
\usepackage{color}
\usepackage{multirow}

\begin{document}

\newcommand{\cbt}{Cr$^{11}$B$_{2}$}
\newcommand{\crb}{CrB$_{2}$}
\newcommand{\mgb}{MgB$_{2}$}
\newcommand{\tbb}{TbB$_{2}$}

\newcommand{\sig}{$\sigma$}
\newcommand{\pei}{$\pi$}

\newcommand{\ozz}{$\langle100\rangle$}
\newcommand{\ooz}{$\langle110\rangle$}
\newcommand{\ooo}{$\langle111\rangle$}
\newcommand{\too}{$\langle211\rangle$}
\newcommand{\red}{\textcolor{red}}

\newcommand{\bpxy}{B-$p_{\text{x,y}}$}

\title{De Haas-van Alphen effect and Fermi surface properties of single crystal CrB$_2$}

\author{M. Brasse}
\affiliation{Lehrstuhl f\"{u}r Physik funktionaler Schichtsysteme,
Technische Universit\"{a}t M\"{u}nchen, D-85748 Garching, Germany}

\author{L. Chioncel}
\affiliation{Theoretical Physics III, Center for Electronic Correlations and Magnetism,
Institute of Physics, University of Augsburg, 86135 Augsburg, Germany}

\author{J. Kune\v{s}}
\affiliation{Institute of Physics, Academy of Sciences, Praha 6 16253, Czech Republic}

\author{A. Bauer}
\affiliation{Physik-Department, Technische Universit\"at M\"unchen,
D-85748 Garching, Germany}

\author{A. Regnat}
\affiliation{Physik-Department, Technische Universit\"at M\"unchen,
D-85748 Garching, Germany}

\author{C.~G.~F. Blum}
\affiliation{Leibniz Institute for Solid State and Materials Research IFW, D-01171 Dresden, Germany}

\author{S. Wurmehl}
\affiliation{Leibniz Institute for Solid State and Materials Research IFW, D-01171 Dresden, Germany}
\affiliation{Institut f\"ur Festk\"orperphysik, Technische Universit\"at Dresden, D-01062 Dresden, Germany}

\author{C. Pfleiderer}
\affiliation{Physik-Department, Technische Universit\"at M\"unchen,
D-85748 Garching, Germany}

\author{M.~A. Wilde}
\email[]{mwilde@ph.tum.de}
\affiliation{Lehrstuhl f\"{u}r Physik funktionaler Schichtsysteme,
Technische Universit\"{a}t M\"{u}nchen, D-85748 Garching, Germany}

\author{D. Grundler}
\affiliation{Lehrstuhl f\"{u}r Physik funktionaler Schichtsysteme,
Technische Universit\"{a}t M\"{u}nchen, D-85748 Garching, Germany}

\date{\today}

\begin{abstract}
We report the angular dependence of three distinct de Haas-van Alphen (dHvA) frequencies of the torque magnetization in the itinerant antiferromagnet {\crb} at temperatures down to 0.3\,K and magnetic fields up to 14\,T. Comparison 
with the Fermi surface calculations considering an incommensurate cycloidal magnetic order suggests that two of the observed dHvA oscillations arise from electron-like Fermi surface sheets formed by bands with strong {\bpxy} character. The third orbit could correspond to a Cr-$d$ derived Fermi surface sheet. The measured effective masses of these Fermi surface sheets display strong enhancements of a factor of about two over the calculated band masses which can be attributed to electron-phonon coupling and electronic correlations. Signatures of further heavy $d$-electron bands that are predicted by the calculations are not observed in the temperature and field range studied. In view that the B-$p$ bands are at the heart of conventional high-temperature superconductivity in the isostructural {\mgb}, we consider possible implications of our findings for nonmagnetic {\crb} and an interplay of itinerant antiferromagnetism with superconductivity.
\end{abstract}

\maketitle

\section{Introduction}

The class of transition-metal and rare-earth diborides $M$B$_2$ ($M=$~Cr, Mn, V, Zr, Nb, Tm, Tb,...), comprises an unusual combination of structural properties and exceptionally large diversity of different electronic ground states\cite{Nagamatsu2001,Barnes1969,Fedorchenko2009,Vajeeston2001,Grechnev2009}. In the hexagonal $C32$ crystal structure closest-packed $M$-layers and honeycomb B-layers alternate along the $\left[001\right]$ direction. An intriguing example for an electronic form of order of general interest in such a layered crystalline environment is the observation of itinerant antiferromagnetism below $T_N=88\,{\rm K}$ in {\crb}, inferred long ago from the resistivity, magnetic susceptibility and specific heat \cite{Barnes1969,Vajeeston2001,Balakrishnan2005}. Yet, relatively little is known about this fascinating material.

First microscopic evidence for itinerant antiferromagnetism in {\crb} was based on NMR studies in powder samples \cite{Kitaoka1978,Kitaoka1980}, which suggested that {\crb} is located in the middle of the local moment and the weakly antiferromagnetic limits. Further evidence for itinerant magnetism was inferred from measurements of the magneto-volume coupling, which revealed that the thermal expansion coefficient in the nonmagnetic state of {\crb} is nearly the same as in the weak itinerant ferromagnet ZrZn$_{2}$ \cite{Nishihara1987}. Moreover, early neutron scattering studies in {\crb} \cite{Funhashi1977} suggested cycloidal magnetic order. An ordering wave vector $\textbf{q} = 0.285\,\textbf{q}_{110}$ and $q_{110} = 2\pi/\frac{a}{2}$ was stated and a reduced ordered magnetic moment of 0.5\,$\mu_{\textrm{B}}\,\textrm{f.u.}^{-1}$ was found, which is characteristic of itinerant magnetism. However, these neutron scattering experiments had to be carried out on thin single crystals to overcome the very strong neutron absorption by the $^{10}$B content of natural boron used for the sample preparation. Recent NMR studies \cite{Michioka2007} extended the information on the magnetic order, suggesting a combination of incommensurate and commensurate spin modulations.

In this paper we report an experimental investigation of the de Haas-van Alphen (dHvA) effect in high-quality single-crystal {\crb} as inferred from quantum oscillatory components of the torque magnetization at temperatures down to 0.3\,K and magnetic fields up to 14\,T. We observed three distinct dHvA frequencies and their characteristic angle dependencies. Temperature dependent data allow us to extract effective masses between 0.86 and 1.22~$m_{\rm e}$, where $m_{\rm e}$ is the free electron mass. The associated mean free paths range from 26 to 69~nm. To analyze our data we performed band structure calculations where we consider a cycloidal magnetic structure of {\crb} below $T_N$. In order to determine the sensitivity of the different Fermi surface sheets to the exact form of the magnetic order we also performed calculations for the collinear ferromagnetic case as well as for the nonmagnetic case. Two of the experimentally observed orbits are in very satisfactory agreement with two electron-like Fermi surface sheets with strong {\bpxy} character. These Fermi surface sheets are shown to be highly insensitive to the magnetic order in the calculations. We therefore  assign the two orbits to the corresponding pockets of the Fermi surface. The large effective masses measured for the corresponding Fermi surface sheets together with the results of the band structure calculations suggest a strong mass enhancement that can arise from electron-phonon coupling and electronic correlations. The third orbit experimentally observed might correspond to a small electron-like Fermi surface sheet with dominant Cr-$d$ character.

In view that MgB$_2$, the conventional superconductor with the largest transition temperature, is an isostructural sibling of {\crb}, our findings motivate a comparison of the electronic structure between these systems. Band structure calculations and quantum oscillatory components of the torque magnetization of {\mgb} suggest consistently \cite{Yelland2002,Carrington2003,Mazin2002}, that two superconducting gaps form on the {\pei}- and {\sig}-sheets of the Fermi surface without any noticeable interband impurity scattering \cite{Kortus2001,Mazin2002a}. In turn, as the {\pei}- and {\sig}-sheets of the Fermi surface arise from the boron orbitals an interesting question concerns the fate of these Fermi surface sheets and the superconducting instability in isostructural transition-metal and rare-earth diborides.  {\crb}, which we address here, may be the perhaps most interesting material in this regard.


Our paper is organized as follows. In section \ref{methods}, we present the experimental and theoretical methods. Experimental results are given in section \ref{results}. This is followed by a detailed presentation of the electronic structure of {\crb} calculated in density functional theory in section \ref{calculation} and a comparison with the experimental results in \ref{FS}. The paper concludes with a short discussion of similarities and differences with {\mgb} and possible implications for the interplay of itinerant antiferromagnetism with superconductivity in the hexagonal diborides.

\section{Experimental and theoretical methods}
\label{methods}

For the dHvA measurements we used a single crystal prepared by optical float-zoning \cite{Bauer:PRB2013}. The feed rods for the float-zoning were made at TU M\"unchen starting from high purity elements, namely 4N5 Cr and B as reacted in a bespoke crucible by radio-frequency induction heating. $99$\% isotopically enriched  $^{11}$B was used to make neutron scattering experiments possible ($^{10}$B is a strong neutron absorber). Following initial tests with an UHV-compatible four-mirror image furnace \cite{Neubauer:RSI2010} the float-zoning was performed with a high pressure image furnace at IFW Dresden \cite{ifw-furnace} to reach the high melting temperature. To suppress the losses due to the vapor pressure of B an inert Ar atmosphere of $15$~bar was applied during growth. Further details of the crystal growth procedure are reported in Ref.~[\onlinecite{Bauer:PRB2013}].\\

Samples were oriented by means of Laue x-ray backscattering and cut using a wire saw. For our dHvA study we used a cuboid of $2.45\times2.2\times0.8$ mm$^{3}$ parallel to $\left[001\right]\times\left[100\right]\times\left[120\right]$, respectively. A series of comprehensive measurements on samples from the same ingot, to be reported in detail elsewhere, establishes an excellent sample quality. The electrical resistivity of our samples decreases monotonically with decreasing temperature and displays a distinct cusp at $T_{\rm N}$ followed by a temperature dependence consistent with a spin-gap of $\sim220\,{\rm K}$. The resistivity is moderately anisotropic by a factor of $1.5$ to $4.5$ at all temperatures (the resistivity for current along $\left[100\right]$ is larger). At $T\rightarrow 0$ the residual resistivity approaches a small sample-dependent value of a few $\mu\Omega{\rm cm}$.\cite{Bauer:PRB2013} The residual resistivity ratio is $11$ for current along $\left[100\right]$ and $31$ for current along $\left[001\right]$. These values are the highest ones reported in literature. Further, the paramagnetic susceptibility is enhanced and displays a Curie-Weiss dependence above $T_{\rm N}$. Most importantly for the work reported in this paper, our crystals of {\crb} do not exhibit the Curie tail observed in previous studies, which excludes magnetic contributions from Fe impurities in contrast to previous reports.\cite{Balakrishnan2005,Castaing1972a} We have performed comprehensive elastic neutron scattering studies showing a complex magnetic order. The results are broadly consistent with the cycloidal magnetic ordering wave vector reported in Ref.~\onlinecite{Funhashi1977}, but await a full refinement.

The dHvA effect was measured by torque magnetometry using cantilevers made of $50$ $\mu$m thick CuBe foils. We mounted the sample onto the mobile plate of the cantilever, which serves as a capacitor plate at the same time. The torque signal $\mathbf{\Gamma}=\mathbf{M}\times\mathbf{B}$ was detected by a capacitive readout scheme using an \textit{Andeen-Hagerling} capacitance bridge.\cite{Wilde2008} The setup had a torque resolution of $\Delta\Gamma\approx 5\cdot 10^{-11}$\,Nm. We performed the measurements in a $^3$He cryostat at temperatures down to $0.3$~K and magnetic fields up to $14$~T. The sample stage was equipped with a mechanical rotator to change the angle of the cantilever with respect to the magnetic field \textit{in situ} with a resolution better than $0.2^{\circ}$.\cite{Rupprecht2013} We investigated the dHvA effect in two principal planes of the hexagonal crystal structure. First, the magnetic field was rotated in the hexagonal basal plane described by the angle $\varphi$ with respect to the $\left[100\right]$ direction [see inset in Fig.~\ref{Fig1}~(b)]. Second, after remounting of the sample on the cantilever rotated by $90^{\circ}$ the field was applied in the $\left[001\right]$-$\left[120\right]$ plane [see inset in Fig.~\ref{Fig2}~(c)]. Here, the angle $\psi$ characterizes the field orientation with respect to the $\left[001\right]$ axis.

We analyzed the dHvA data using the Lifshitz-Onsager relation and the Lifshitz-Kosevich formalism following Refs.~[\onlinecite{Shoenberg1984,Wasserman1996}]. Further, we performed electronic structure calculations in the framework of the local spin density approximation (LSDA) method of density functional theory by using the full-potential linearized augmented plane-wave method implemented in the WIEN2k package.\cite{Wien2K,Laskowski2004,Kunes2004} In the calculations the experimental values for the lattice parameters $a=2.969$~\r{A} and $c=3.066$~\r{A} were used. The results were compared to calculations in the Generalized Gradient Approximation (GGA) of the exchange and correlation functionals and were found to be consistent. We have studied non-spin-polarized, collinear spin polarized and cycloidal spin polarized solutions. Cycloidal magnetic order with ordering vector $\textbf{q}$ was treated using the generalized Bloch theorem. \cite{Sandratskii1998} Comparing the total energies for several $\textbf{q}$-vectors along the $\left[110\right]$ and $\left[100\right]$ directions the lowest energy was found for $\textbf{q} = 0.3\pm0.05\,\textbf{q}_{110}$ in good agreement with $\textbf{q} = 0.285\,\textbf{q}_{110}$ found in Ref.~\onlinecite{Funhashi1977} and substantiated by our neutron scattering experiments. Calculations for the collinear ferromagnetic state and the nonmagnetic state were performed in order to find out how sensitive different parts of the Fermi surface are to the magnetic order. In the present study we focus on the Fermi surface properties at the ordering vector $\textbf{q} = 0.3\,\textbf{q}_{110}$. A detailed investigation of the total energy versus $\textbf{q}$ landscape will be the subject of a future study. DHvA frequencies and band masses $m_{\text{b}}$ were extracted from the band structure calculations using the SKEAF (Supercell K-space Extremal Area Finder)\cite{Rourke2012} tool.

\section{Experimental Results}
\label{results}

\begin{figure}[htbp]
\center
\includegraphics[width=8cm]{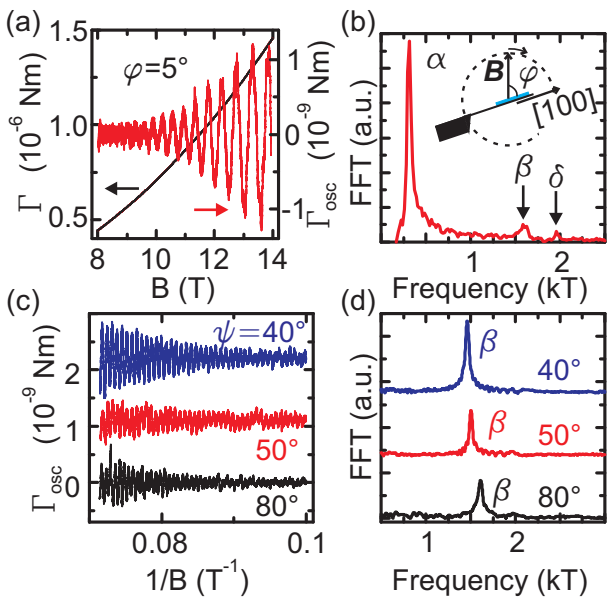}
\caption{(Color online) (a) The torque $\Gamma$ (black line) and the oscillatory torque $\Gamma_{\text{osc}}$ (light line) as a function of magnetic field $B$ for $\varphi=5^{\circ}$ at $T=0.3$ K. (b) Fourier Transform of $\Gamma_{\text{osc}}(1/B)$ revealing three distinct orbits $\alpha$, $\beta$ and $\delta$. A sketch of the experimental geometry is depicted in the inset. (c) $\Gamma_{\text{osc}}$ for three different angles $\psi$ at $T=0.3$ K. (d) Corresponding FFTs showing one dHvA orbit that shifts in frequency.}
\label{Fig1}
\end{figure}

Figure \ref{Fig1} (a) shows the torque data as measured versus magnetic field (black line) in the hexagonal plane for $\varphi=5^{\circ}$ and at $T=0.3$ K.
The quantum oscillations are superimposed on a monotonic background signal. In order to extract the oscillatory signal component $\Gamma_{\text{osc}}$ induced by the dHvA effect a high-order polynomial was fitted and the background subtracted. The dHvA frequencies $f_{i}$ were then determined by Fast-Fourier-Transforms (FFT) of $\Gamma_{\text{osc}}(1/B)$. The FFT spectrum for $\varphi=5^{\circ}$ is shown in Fig. \ref{Fig1} (b). We identify three distinct frequencies, $f_{\alpha}$, $f_{\beta}$, and $f_{\delta}$ that we will attribute to different extremal orbits below. The cross-sectional area $P_i$ of an extremal orbit is, according to the Lifshitz-Onsager relation \cite{Onsager1952}, $P_i=2\pi ef_i/\hbar$. In Fig. \ref{Fig1} (c) and (d) the field was applied in the $\left[001\right]$-$\left[120\right]$ plane. Here, only one frequency $f_{\beta}$ is clearly resolved.

\begin{figure}[htbp]
\center
\includegraphics[width=8cm]{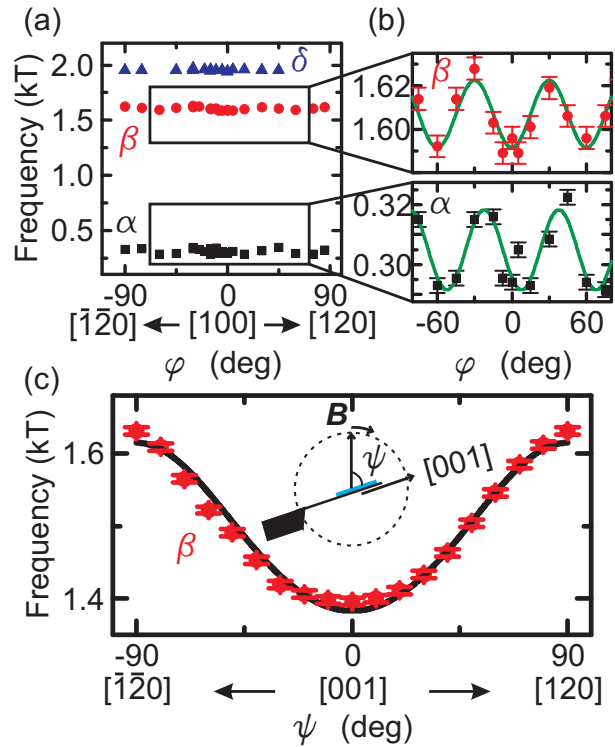}
\caption{(Color online) (a) Angular dependence of the dHvA frequencies for a rotation of the magnetic field in the basal plane showing three orbits $\alpha$, $\beta$ and $\delta$. (b) Expanded view of the angular dependence of $\alpha$ and $\beta$. Both orbits exhibit an apparent $60^{\circ}$ periodicity. Lines are guides to the eyes. (c) Rotation of the magnetic field in the $\left[001\right]$-$\left[120\right]$ plane. The middle frequency branch $\beta$ is present, whereas $\alpha$ and $\delta$ are missing. The data is fitted by a curve (solid line) expected for an ellipsoidal Fermi surface sheet (see text). The inset depicts a sketch of the experimental geometry.}
\label{Fig2}
\end{figure}

In both principal planes the angular dependence of the dHvA effect was studied. A map of both directions is presented in Fig.~\ref{Fig2}.
When rotating the field within the basal plane of the crystal [see Fig.~\ref{Fig2} (a)], three orbits $\alpha$, $\beta$ and $\delta$ with $f_{\alpha}\approx 300$ T, $f_{\beta}\approx 1600$ T and $f_{\delta}\approx 1950$ T, respectively, are identified. They were traced over an angular regime $\Delta \varphi>90^{\circ}$, indicating that the orbits belong to closed Fermi surface sheets. Further, they show only a small variation in frequency indicating that the cross sectional area perpendicular to the basal plane varies only slightly as a function of $\varphi$. In Fig.~\ref{Fig2}~(b) an expanded view of the frequency branches $\alpha$ and $\beta$ is shown. The data exhibit apparently a small but clear $60^{\circ}$ periodicity. The variation corresponds to a peak-to-peak change in extremal area of about $10$~\% for orbit $\alpha$ and $2$~\% for orbit $\beta$. We do not make a statement about the periodicity for orbit $\delta$ since it was not resolved over the whole angular regime. Whether this is an intrinsic effect or due to experimental limitations is unclear at present.

The angular dependence of the dHvA frequencies in the perpendicular plane is shown in Fig.~\ref{Fig2}~(c). Only the $\beta$ orbit is present. The orbits $\alpha$ and $\delta$ are not resolved. This second data set shows that the $\beta$ orbit can be traced in both symmetry planes, again identifying the Fermi surface corresponding to $\beta$ as a closed surface. In a first approach, we describe this Fermi surface as an elongated ellipsoid. The orbit of the extremal area $P_{\beta}$ is then the intersection of a plane with an ellipsoid \cite{Schneider2012}, given by $P_{\beta}=\pi vw/\sqrt{\sin^2{\psi}+\left(v^2/w^2\right)\cos^2{\psi}}$, where $v$ and $w$ denote the semi-major and semi-minor axes of the ellipse in reciprocal space. The resulting fit shown in Fig.~\ref{Fig2}~(c) is in good agreement with the data points and yields $v=\left(2.38\pm 0.01 \right)\cdot 10^9$~m$^{-1}$ and $w=\left(2.05\pm 0.01 \right)\cdot 10^9$~m$^{-1}$. Here, $v$ is the reciprocal-space extent of the ellipsoid along the $\left[001\right]$ direction. We thus deal with a Fermi surface sheet that is well described as an ellipsoid slightly elongated in the $\left[001\right]$ direction and has a very small six-fold modulation of its surface in the basal plane.

The absence of orbits $\alpha$ and $\delta$ for $B$ applied in the $\left[001\right]$-$\left[120\right]$ plane cannot be explained by open Fermi surface sheets as already inferred above. In addition, when we compare both data sets [Fig.~\ref{Fig2}~(a) and (c)] for $B$ aligned along the symmetry direction $\left[\overline{1}\overline{2}0\right]$, the frequencies are visible in the experiment for one direction of the torque while they are absent for the other. If an orbit turns into an open orbit in a certain field direction, this will cause its disappearance regardless of the torque direction. Thus, the disappearance of this orbit in only one torque direction cannot be caused by the orbit turning into an open orbit, but must be caused by another mechanism. Following Ref.~\onlinecite{Shoenberg1984}, the torque for $B$ applied in the $\left[001\right]$-$\left[120\right]$ plane can be written as $\Gamma_{\text{osc}}=-\frac{1}{f}\frac{\partial f}{\partial\psi}M_{\parallel}B$, where $f$ denotes the dHvA frequency and $M_{\parallel}$ the magnetization component parallel to the magnetic field. For a Fermi surface with low anisotropy $\frac{\partial f}{\partial\psi}$ can go to zero \cite{Shoenberg1984}, causing the oscillatory torque signal to vanish in specific symmetry planes. This effect could account for the absence of the frequency branches $\alpha$ and $\delta$ in Fig.~\ref{Fig2}~(c).

We further analyze the dHvA data and extract the effective masses $m^*$ and scattering times $\tau$ from the standard Lifshitz-Kosevich (LK) formalism\cite{Shoenberg1984,Wasserman1996}. The first harmonic of the oscillatory torque is then given by
\begin{equation}
\Gamma_{\text{osc}}\propto \sum_{\text{orbits}} B^{3/2}R_{\text{D}}R_{\text{T}}R_{\text{S}}\sin\left(\frac{2\pi f}{B}+\gamma\right),
\label{eq:formel1}
\end{equation}
where $\gamma$ is the phase, $R_{\text{D}}=\exp(-\pi m_{\text{b}}/eB\tau)$ is the Dingle factor accounting for impurity scattering, $m_{\text{b}}$ is the electronic band mass and $\tau$ is the scattering time. Thermal damping of the oscillations is included in terms of the factor $R_{\text{T}}=X/\sinh X$, where $X=(\frac{2\pi^2k_{\text{B}}m^*T}{\hbar eB})$. Here, $m^*$ is the quasi-particle effective mass which is renormalized by electron-phonon and electron-electron interactions. We set the spin-splitting factor $R_{\text{S}}=1$.
\begin{figure}[htbp]
\center
\includegraphics[width=8cm]{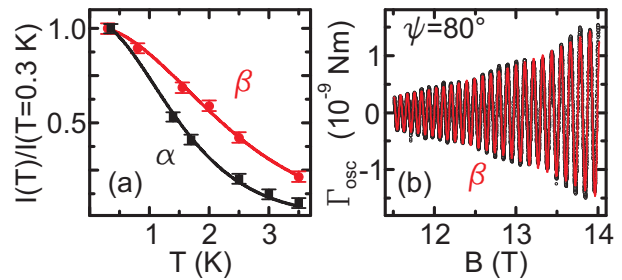}
\caption{(Color online) (a) Normalized temperature-dependent FFT amplitudes $I(T)$ (symbols) and LK fit (solid line) as a function of temperature. The FFT amplitudes were obtained using a fixed-field window. In the expression for $R_{\rm T}$ the average field of this window was used. (b) Torque vs. $B$-field at $T=0.3$ K (symbols) and fit (solid line) of Eq. (\ref{eq:formel1}) using $m^*=0.86~{\rm m}_{\text{e}}$ as determined above.}
\label{Fig3}
\end{figure} 
\begin{table*}[htbp]
	\centering
		\begin{tabular}{l|l|l|l|l|l|l|l}
			Band & Orbit & $f_{\text{calc}}$ (T) & $f_{\text{exp}}$ (T) & $m^*/m_{\text{e}}$ & $m_{\text{b}}/m_{\text{e}}$ & $l$ (nm) & $\lambda$ \\
			\hline
			\hline
			Cr-$d$ & $\alpha$ & $(734)$ & $308$ & $1.22\pm0.12$ & $(0.93)$ & $26$ & $(0.3)$\\
			\hline
			\multirow{2}{10mm}{B-$p$}
			& $\beta$ & $1899$ & $1608$ & $0.86\pm0.07$ & $0.37$ & $69$ & $1.3$ \\
      & $\delta$ & $2452$ & $1951$ & $1.07\pm0.06$ & $0.53$ & $67$ & $1.0$ \\
			\hline
		\end{tabular}
	\caption{Experimental and calculated dHvA frequencies, effective masses $m^*$, band masses $m_{\text{b}}$,  mean free paths $l$ and electron-phonon coupling constants $\lambda$. The calculated values for orbit $\alpha$ are given in brackets because they refer to the dumbbell-shaped pocket which cannot be assigned unambiguously.}
	\label{tab:table1}
\end{table*}

The temperature dependence of the dHvA effect for orbits $\alpha$ and $\beta$ is shown in Fig.~\ref{Fig3}~(a). We applied the magnetic field $B$ at small angles off the symmetry axes, at $\varphi=5^{\circ}$ and $\psi=80^{\circ}$, respectively, since the torque signals vanish for the field parallel to the symmetry axes. To determine the effective masses $m^*$, we plotted the amplitudes of the FFT peaks as a function of $T$. We then fitted the expression for $R_{\text{T}}$ to the data, substituting the average field of the FFT window $\overline{B}=(B_{\text{min}}+B_{\text{max}})/2$ for $B$. The effective masses determined from the LK fits are $m^*_{\alpha}/m_{\text{e}}=1.22\pm0.12$ and $m^*_{\beta}/m_{\text{e}}=0.86\pm0.07$ (Table~\ref{tab:table1}). The error bars were determined from the standard deviation of the LK fit. For orbit $\delta$, the LK formalism was applied at temperatures $T\leq1.5$ K due to the small signal-to-noise ratio at larger $T$. We obtain an effective mass $m^*_{\delta}/m_{\text{e}}=1.07\pm0.06$ for this orbit. This value might contain a systematic error, since only few data points could be collected.

The mean free paths are determined by fitting Eq. (\ref{eq:formel1}) to our data as shown in Fig.~\ref{Fig3}~(b). We used the values of $m^*$  listed in Table~\ref{tab:table1} as input parameters and obtained the Dingle factor $R_{\text{D}}$. Following Refs.~[\onlinecite{Yelland2002,Arnold2011}], the mean free paths are derived as follows: For free electrons we consider $m_{\text{b}}/\tau=\hbar k_{\text{F}}/l$, where $k_{\text{F}}$ denotes the Fermi wave vector and $l$ is the mean free path. $k_{\text{F}}$ is then replaced by $\pi k^2_{\text{F}}=2\pi ef/\hbar$ which amounts to the approximation that the frequency $f$ arises from a circular area in reciprocal space. Values for the mean free path extracted this way are between $26$ and $69$~nm (Table~\ref{tab:table1}) depending on the orbit.

\section{Discussion}

\subsection{Description of the calculated electronic structure}
\label{calculation}

\begin{figure*}[htbp]
\center
\includegraphics[width=14cm]{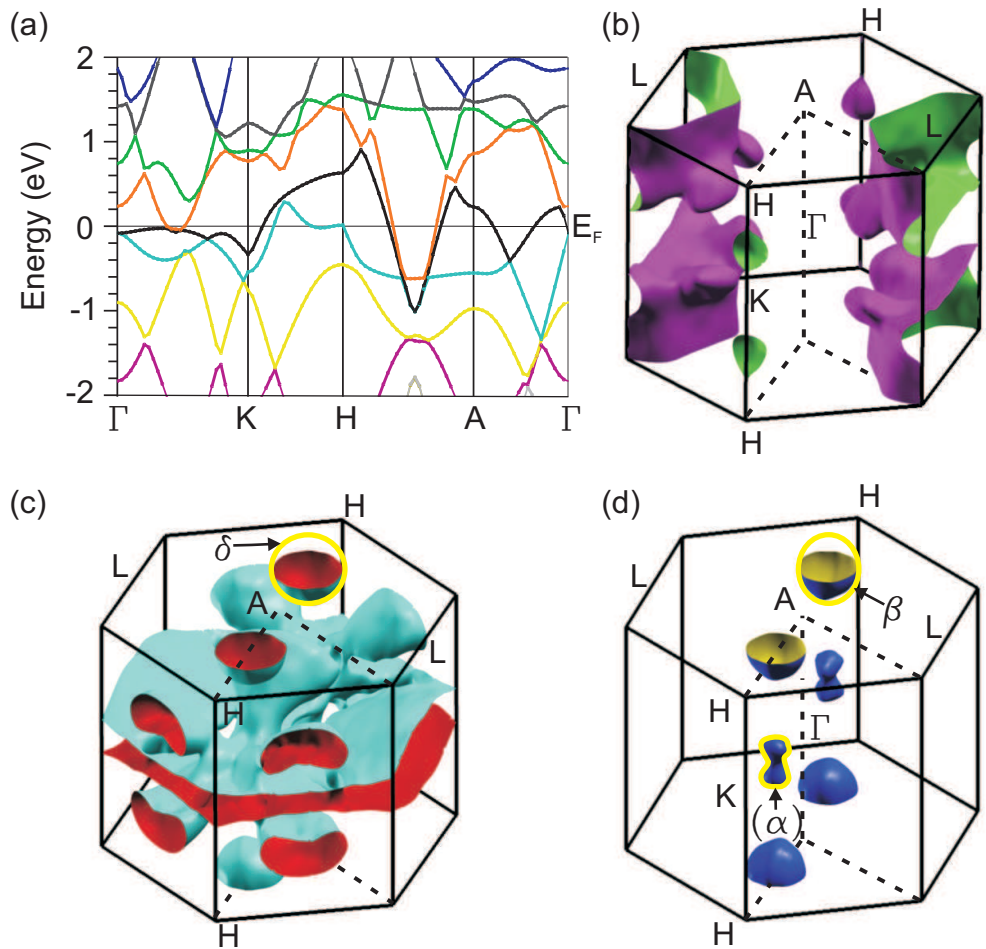}
\caption{ (Color) Calculated electronic structure of {\crb} with a cycloidal magnetic ordering wave vector $\textbf{q} = 0.3\,\textbf{q}_{110}$ using the WIEN2k package. (a) Band structure of {\crb} around the Fermi level $E_{\textrm{F}}$. In total, three bands (blue, orange, black) cross $E_{\textrm{F}}$. (b-d) Fermi surface sheets of {\crb} corresponding to the three bands that cross $E_{\textrm{F}}$ in (a). For clarity, the Fermi surfaces sheets corresponding to the three different bands are plotted separately. (b) The Fermi surface sheet corresponding to the blue band consists of a singly connected structure centered around the L point at the boundary of the 1. Brillouin zone. It has predominantly Cr-$d$ character. (c) Fermi surface corresponding to the black band. Two ball-shaped, B-$p$ derived Fermi surface pockets are present between the A- and the H-points. The extremal orbit $\delta$ assigned to the experimentally observed dHvA frequency is shown in yellow. The ball-shape of this pocket is quite insensitive to magnetic order (see main text). The remaining Fermi surface sheets of this band exhibit a complicated structure and are of Cr-$d$ character. Thorough analysis with SKEAF showed that none of the numerous further extremal orbits of these sheets are close to the experimental ones. (d) Two ball-shaped and two dumbbell-shaped Fermi surface pockets originate from the orange band in (a). The orbit $\beta$ is allocated to the ball-shaped pocket as indicated. This pocket is also predominantly B-$p$-like and very insensitive to the magnetic order. The dumbbell-shaped pockets have Cr-$d$ orbital character. The frequency of the $\alpha$ orbit seen in the experiment matches reasonably well to this pocket (see Table~\ref{tab:table1}), while the angular dependence does not. In light of the fact that the pocket is of Cr-$d$ orbital character and thus very sensitive to the exact form of the magnetic order we refrain from a clear assignment.}
\label{Fig4}
\end{figure*}

The calculated electronic structure of {\crb} assuming cycloidal magnetic order as described in section \ref{methods} is depicted in Fig.~\ref{Fig4}.
 From the calculations we infer that mainly B-$p$ and Cr-$d$ states are present at the Fermi level. Figure~\ref{Fig4}~(a) shows the band structure. In total, three bands (denoted in blue, black and orange color) cross the Fermi level. The corresponding Fermi surfaces are shown in Figs.~\ref{Fig4}~(b-d). Fermi surface sheets originating from different bands are plotted separately for clarity. The Fermi surface shown in (b) corresponds to the band depicted in blue and consists of a singly connected trunk-like structure with multiple extrusions centered around the $L$ points at the Brillouin zone boundary. It has dominantly Cr-$d$ character and is hole-like. In Fig.~\ref{Fig4}~(c) the Fermi surface sheets corresponding to the black band in (a) are depicted. Here a complicated multiply connected structure centered around $\Gamma$ and multiple copies of two different singly connected pockets are present. It is important to note that in this band only the two copies of the pocket located between the A- and H- point have a ball-like closed surface of high symmetry, like that observed in the experiment. Except for this electron pocket, which is derived from B-$p$ states, the Fermi surface sheets of this band can be traced to Cr-$d$ orbitals. In Fig.~\ref{Fig4}~(d) we show the Fermi surface pockets arising from the band depicted in orange. Here, two copies of a ball-shaped pocket located between A and H and two copies of a dumbbell-shaped pocket are present. They are closed and electron-like. The ball-shaped pocket is due to B-$p$ orbitals whereas the dumbbell-shaped sheet has dominantly Cr-$d$ character. We note that the ball-shaped pockets in (c) and (d) also occur in the nonmagnetic calculation\cite{supmaterial} as single copies centered around the A point. The doubling of the pockets and the shift along the A-H direction in the magnetic calculation is a direct consequence of the lifting of the spin degeneracy and the direction of the magnetic ordering wave vector, respectively.

The calculations suggest an ordered magnetic moment of 1.3\,$\mu_{\textrm{B}}\,\textrm{f.u.}^{-1}$, while early experiments\cite{Funhashi1977} reported $~0.5\mu_{\textrm{B}}$. In light of the fact that our neutron scattering data as well as the NMR experiments in Ref.~\onlinecite{Michioka2007} show hints of further ordering vectors, the resolution of this discrepancy between experiment and theory is beyond the scope of the present paper. Since this uncertainty could potentially influence the interpretation of our dHvA data we performed additional band structure calculations to assess to what extent the Fermi surface sheets we associate with the experimental orbits are sensitive to the magnetic order. In particular we performed the calculations for the collinear spin state and for nonmagnetic {\crb} \cite{supmaterial}. The general outcome is, that the two different ball-shaped Fermi surface sheets with B-$p$ orbital character are quite insensitive to the magnetic order (see Supplemental Material \cite{supmaterial}). The main effect in going from the nonmagnetic to the cycloidal order is the lifting of the spin degeneracy and the corresponding shift of the pockets away from the A point along the A-H direction. The changes in shape and size are only minor. In contrast, the remaining Fermi surfaces sheets, which are dominantly of Cr-$d$ orbital character, change radically between the cycloidal magnetic, collinear magnetic and non-magnetic calculations, i.e., they are highly sensitive to the type and magnitude of the magnetic order parameter.

\subsection{Comparison with Experiment}
\label{FS}

The extremal-orbit dHvA frequencies and the band masses $m_{\text{b}}$ of the calculated Fermi surface with cycloidal magnetic order were analyzed with SKEAF\cite{Rourke2012}. A large number of extremal orbits was found due to the complicated structure of the Fermi surface in Fig.~\ref{Fig4}. To allocate the three experimentally observed orbits $\alpha$, $\beta$ and $\delta$, we compared them with the calculated Fermi surface considering the following criteria: the dHvA frequency and its angular dependence (giving information about size, shape and topology), the charge carrier masses and last but not least the exclusiveness (i.e., are there other candidates that might also match, or is the assignment unique?).

Following these criteria, the experimentally observed orbits $\beta$ and $\delta$ were assigned to the two ball-shaped sheets between A and H formed by B-$p$ electrons as listed in Table~\ref{tab:table1} and illustrated in Fig.~\ref{Fig4}~(c-d). This assignment can be made due to (i) the good frequency (Fermi surface cross section) match, (ii) the closed-surface topology with nearly spherical shape in both experiment and calculation, (iii) the sufficiently light band masses and (iv) the fact that there are no other Fermi surface sheets that are anywhere close to match these criteria. In particular, the Cr-$d$ derived sheets look very different, such that even moderate changes in the magnetic order are unlikely to produce similar pockets. In addition, the charge carriers of most Cr-$d$ derived sheets are much too heavy.

The measured frequencies differ by $300-500$~T from the calculations. This corresponds to a mismatch in $k$-space extent of only $0.2$ to $0.3$~\% of the Brillouin zone cross sectional area. This is thus is a very satisfactory match. Rigid band shifts on the order of $100$~meV can bring the calculated frequencies of $\beta$ and $\delta$ into coincidence with the experimental results. \cite{Carrington2007} However, the shape of the remaining dumbbell-shaped, Cr-$d$ derived, Fermi surface pockets is heavily affected by the band shift.

Apart from the absolute frequencies, several experimentally observed details of the Fermi surface shape are reproduced qualitatively in the calculations for the cycloidal order: The nearly ball-shaped pocket giving rise to the $\beta$ orbit is slightly elongated in the $[001]$ direction in the experiment as well as in the calculation. The very small sixfold modulation of the $\beta$ orbit for rotation in the basal plane seen in the experiment [Fig.~\ref{Fig2}~(b)] is already present in the nonmagnetic calculation. In the cycloidal magnetic calculation, performed for one single domain of the ordering vector, the weak modulation is also present, but with reduced symmetry. However, assuming that the different domains corresponding to the symmetry related $\textbf{q}$-vectors are equally populated, an apparent sixfold symmetry is restored, provided that the overlapping frequency branches from the different domains are not spectrally resolved in the FFT. For the observed frequency modulation amplitude of $~15$~T this condition is fulfilled in the experiment.

The measured frequency and topology of orbit $\alpha$ matches best with the dumbbell-shaped electron pocket along $\Gamma$-$K$ shown in Fig.~\ref{Fig4}~(d). It is therefore tempting to \emph{tentatively} assign $\alpha$ to this Fermi surface sheet as indicated with $(\alpha)$ in Fig.~\ref{Fig4}~(d) and Table~\ref{tab:table1}. The frequency match is satisfactory, it is a closed surface  and there is no other sheet providing similarly small cross sections on a closed surface. The measured effective mass $m^*/m_{\textrm{e}}=1.22 \pm0.12$ and the band mass $m_{\textrm{b}}/m_{\textrm{e}}=0.93$ are in good agreement. However, that the dHvA frequency vanishes in the experiment upon field rotation in the $\psi$-direction cannot be explained by a vanishing torque due to $\frac{\partial f}{\partial \psi} \rightarrow 0$ for the dumbbell structure. On the contrary, a signal from the neck orbit of the dumbbell would be expected to appear upon rotation about $\psi$. As noted above, the dumbbell electron pocket has dominantly Cr-$d$ character and its cross section and topology is thus heavily affected by the magnetic order. In account of these doubts we refrain from a definite assignment of the $\alpha$ orbit to the dumbbell pocket.

Table~\ref{tab:table1} summarizes the calculated as well as the measured orbits for the field applied at an angle $\varphi=5^{\circ}$ off the $\left[100\right]$-axis, since the torque becomes small for $B||[100]$. The calculated values for the $\alpha$ orbit are given in brackets as they refer to the dumbbell pocket that cannot clearly be assigned.

The calculated Fermi surface of {\crb} suggests many more possible extremal orbits, all arising from Cr$-d$ dominated Fermi surface parts. In most cases, these are large orbits with heavy effective masses, such that lower temperatures and higher magnetic fields might enable their observation. This would in turn allow to refine the present picture of the Cr-$d$ Fermi surface sheets and, consequently, the details of the magnetic order.

Comparing the calculated band masses of the charge carriers with the experiment, we find that the quasiparticle masses $m^*$ are strongly enhanced over the band masses $m_{\text{b}}$ for the B-$p$ pockets (Table~\ref{tab:table1}). We attribute this finding to many-body interactions that are not included in the band structure calculations.\cite{Wasserman1996} Assuming that electron-phonon interaction is the dominant source of these interactions in the B-$p$ derived Fermi surface sheets as it is in the isostructural compound MgB$_2$, we can calculate an upper bound for the electron-phonon coupling constant $\lambda$ defined by $m^*=(1+\lambda)m_{\text{b}}$.\cite{Carrington2003} For the orbits $\beta$ and $\delta$ the values of $\lambda$ are $1.3$ and $1.0$, respectively (Table~\ref{tab:table1}). Considering the $\sigma$-bonding $p_{\text{x,y}}$ character that suggests strong bonds being sensitive to the B-B bondlength, a strong electron-phonon coupling in the B-$p$ sheets of {\crb} seems likely. An analogous discussion has been presented for MgB$_2$ as we will detail in the next section.

\subsection{Comparison with {\mgb}}

In MgB$_2$ two-band superconductivity arises from the $\sigma$ and $\pi$ bands formed by the {\bpxy} and by the B-$p_{\text{z}}$ electrons, respectively. Especially the former show large electron-phonon coupling constants of $\lambda=0.96-1.2$ \cite{Yelland2002,Liu2001} and are responsible for the high transition temperature. The strength of this electron-phonon coupling is directly related to having partially filled $\sigma$-bonding {\bpxy} states at the Fermi level: Since $\sigma$ bonds are strong and the {\bpxy} states are sensitive to the B-B bondlength this leads to a large electron-phonon coupling. A further key ingredient for the superconductivity is the rather weak inter-band scattering.

Our dHvA experiments and density functional theory calculations show that parts of the Fermi surface in the isostructural itinerant antiferromagnet {\crb} also derive from bands with clear $p_{\text{x,y}}$ character \cite{supmaterial}. Furthermore, we find putative evidence for large electron-phonon coupling constants for these Fermi surface sheets with upper bounds of $\lambda=1.0-1.3$ that compare well to the values in MgB$_2$. However, in sharp contrast the {\bpxy}-electron derived Fermi surface sheets in {\crb} are closed three-dimensional sheets as opposed to the cylindrical two-dimensional open Fermi surface sheets in {\mgb} \cite{Carrington2003}.

It is now interesting to speculate on routes that may lead to superconducting instabilities in {\crb}. First, in the absence of itinerant antiferromagnetism and ignoring for the moment the presence of the $d$-electron bands, the electron-phonon coupling may be sufficient to stabilize conventional superconductivity. Based on the differences of the topology of the $p$-derived bands the superconducting transition temperature may thereby be much lower. However, in preliminary calculations we have found that the closed Fermi surface sheets partly expand when the $c/a$ ratio is reduced. The superconducting transition temperature of nonmagnetic {\crb} may in turn increase. This suggests that uniaxial stress may represent a control parameter to tune a superconducting instability in nonmagnetic {\crb}. Second, if there is only weak inter-band scattering, antiferromagnetic {\crb} may potentially display a coexistence of antiferromagnetism and superconductivity, where the $p$-derived bands undergo a superconducting instability, while the $d$-derived bands support antiferromagnetic order. Third, located between these two scenarios a constructive interplay of antiferromagnetism and superconductivity may occur in nonmagnetic {\crb} precisely at the border of a zero temperature instability of the itinerant antiferromagnetism. This scenario of an antiferromagnetic quantum critical point has been considered in a wide range of materials.\cite{Ueda2003} Tuning {\crb} towards such a point may allow to study in a controlled manner the evolution from conventional electron-phonon mediated superconductivity to itinerant antiferromagnetism. This way one of the perhaps most pressing hidden agendas in this field, notably a constructive role of magneto-elastic coupling in the formation of very high superconducting transition temperatures, may be accessed.

\section{Conclusions}

In summary, we have studied the angular dependence of the dHvA effect in a single crystal of CrB$_2$ at low temperatures by means of cantilever torque magnetometry. We have observed three extremal orbits, all belonging to closed Fermi surface sheets. Two have a small apparently sixfold surface modulation in the basal plane. The temperature dependence of the dHvA effect provides the orbits effective masses ranging from 0.8 to 1.22~$m_{\rm e}$. Band structure calculations have been performed for the cycloidal magnetic order, the collinear ferromagnet and the nonmagnetic state. We allocated two orbits to B-$p$-derived Fermi surface pockets of nearly spherical geometry. The pockets are shown to be insensitive to the magnetic order. The comparison of the measured effective masses $m^*$ and with the band structure values $m_{\rm b}$ from the calculation reveals an upper bound of the electron-phonon coupling constant of $1-1.3$. This value is comparable to the electron-phonon coupling constant in the corresponding B-$p$ sheets in the isostructural MgB$_2$, which are responsible for its high superconducting transition temperature. This analogy makes {\crb} a candidate to explore possible routes to superconductivity when tuning the electronic structure of {\crb} with clean tuning parameters such as hydrostatic and/or uniaxial pressure.

We thank G. Behr, B. B\"uchner, A. Erb, C. Franz, M. Halder, W. L\"oser, W. Kreuzpaintner, S. Mayr, M. Meven, C. Morkel, B. Pedersen, M. Schulze, A. Senyshin, M. Wagner as well as the research technology department of the IFW Dresden for fruitful discussions and technical support. AB, AR and MB acknowledge support through the TUM Graduate School. SW acknowledges support by the Deutsche Forschungsgemeinschaft (DFG) under the Emmy-Noether program in project WU595/3-1. Financial support through DFG TRR80 is gratefully acknowledged.

\end{document}